\documentclass[prd,a4paper,showpacs]{revtex4}
\usepackage{epsfig}
\usepackage{color}
\usepackage{amssymb}
\usepackage{amsfonts}
\usepackage{graphicx}
\usepackage{keyval,graphicx}
\usepackage{textcomp,wasysym}

\begin{document}
\title{Updated study of the $\eta_c$ and $\eta_c^\prime$ decays into light vector mesons }
\author{Qian Wang$^1$, Xiao-Hai Liu$^1$, Qiang Zhao$^{1,2}$}

\affiliation{1) Institute of High Energy Physics, Chinese Academy of
Sciences, Beijing 100049, P.R. China \\
2) Theoretical Physics Center for Science Facilities, CAS, Beijing
100049, China}

\begin{abstract}

We re-investigate the exclusive decays of $\eta_c$ and
$\eta_c^\prime$ to a pair of light vector mesons, i.e. $\eta_c
(\eta_c')\to VV$. The long-distance intermediate meson loop (IML)
effects are evaluated as a non-perturbative mechanism in addition to
the short-distance $c\bar{c}$ annihilation contributions. We show
that both processes can be reasonably well constrained with the help
of the available experimental data. Since $\eta_c$ and $\eta_c'$ are
the spin-0 partners of $J/\psi$ and $\psi'$, respectively, our study
is useful for gaining insights into the pQCD helicity selection rule
violations in charmonium decays and the long-standing ``$\rho\pi$
puzzle".

\end{abstract}

\date{\today}

\maketitle

\section{Introduction}

Our knowledge about the properties of $\eta_c$ and $\eta_c'$ is
still limited. Although $\eta_c'$ as the first radial excitation
state of the $\eta_c$ has been predicted and studied for a long
time, it is very recently that we gained more detailed information
about its decay properties in experiment at CLEO~\cite{:2009vg} and
BES-III~\cite{BESIII:2011ac}. The exclusive decays of $\eta_c$ and
$\eta_c^\prime$ to a pair of light vector mesons are of great
interests among many other properties of these two states. One
reason is that the $VV$ decay channel turns out to be one of the
most important decay channels for both $\eta_c$ and $\eta_c'$. For
instance, the branching ratios for $\eta_c\to VV$ are typically at
order of $10^{-3}$ to $10^{-2}$. In contrast, these decay channels
should be suppressed by the so-called helicity selection rule
(HSR)~\cite{Brodsky:1981kj,Chernyak:1981zz,Chernyak:1983ej}. Such an
observation of the HSR violation indicates the importance of QCD
higher twist contributions or presence of a non-pQCD mechanism that
violates the HSR. The contradiction between the data and the HSR
expectations based on the perturbative QCD (pQCD) has drawn much
attention, and various attempts have been made to understand the
underlying
dynamics~\cite{Benayoun:1990ey,Anselmino:1990vs,Zhou:2005fc,Anselmino:1993yg,Zhao:2006cx,Braaten:2000cm,Santorelli:2007xg,Sun:2010qx}.

Another reason is that $\eta_c(\eta_c^\prime)$ exclusive decays can
also be closely related to the long-standing so-called ``$\rho\pi$
puzzle" in $J/\psi (\psi')\to VP$. The decays of $J/\psi$ and
$\psi'$   into light hadrons are supposed to be via the valence
$c\bar{c}$ annihilations into three gluons in pQCD to the leading
order at a typical distance of $1/m_c$. Thus, the following relation
similar to that between $J/\psi$ and $\psi'$ can be expected:
\begin{eqnarray}\label{rela-1}
R_{\eta_c\eta_c'} \equiv \frac{BR(\eta_c'\to 2g)}{BR(\eta_c \to 2g)}
= \frac{BR(\eta_c'\to \gamma\gamma)}{BR(\eta_c \to \gamma\gamma)}
 .
\end{eqnarray}
In the heavy quark limit, i.e. $m_c$ is infinitely large, the mass
difference between $\eta_c$ and $\eta_c'$ can be neglected. Hence,
the branching ratio fraction becomes
\begin{eqnarray}
R_{\eta_c\eta_c'}  \simeq \left|\frac{\eta_c'(0)}{\eta_c(0)}
\right|^2 \frac{\Gamma_{tot}^{\eta_c}}{\Gamma_{tot}^{\eta_c'}} \ ,
\end{eqnarray}
similar to the ``12\% rule" between the $J/\psi$ and $\psi'$ decays.
In the above equation $\eta_c(0)$ and $\eta_c'(0)$ are the values of
the $\eta_c$ and $\eta_c'$ wavefunctions at their origins,
respectively. On the other hand, in terms of the leading
contributions from a potential quark model, deviations can be
expected and one has
\begin{eqnarray}\label{rela-2}
R_{\eta_c\eta_c'} =\left(\frac{M_{\eta_c}}{M_{\eta_c'}}\right)^2
\left|\frac{\eta_c'(0)}{\eta_c(0)} \right|^2
\frac{\Gamma_{tot}^{\eta_c}}{\Gamma_{tot}^{\eta_c'}} \ ,
\end{eqnarray}
where the kinematic corrections cannot be neglected.

In the case of the $J/\psi$ and $\psi'$ decays, many theoretical
efforts have been made in order to understand the origin of a
significant deviation from the ``12\% rule" in $J/\psi (\psi')\to
\rho\pi$ and $K^*\bar{K}+c.c.$ In recent
works~\cite{Zhao:2006gw,Li:2007ky,Wang:2012mf}, we show that the
interferences between the strong and EM decay amplitudes in both
$J/\psi (\psi')\to VP$ are essential for understanding the
``$\rho\pi$ puzzle". Similar ideas had been proposed in the
literature~\cite{Suzuki:2001fs,Seiden:1988rr}. A numerical study of
the overall decay channels for $J/\psi (\psi')\to VP$ indeed
suggests such a phenomenon~\cite{Li:2007ky}. In
Ref.~\cite{Wang:2012mf}, it is shown that the intermediate charmed
meson loops can be considered as a long-distance effect to suppress
the strong transition amplitudes due to their destructive
interference with the short-distance strong transition amplitude in
the $\psi^\prime$ decays.

\begin{figure}
\begin{center}
\hspace{-4cm}
\includegraphics[scale=0.5]{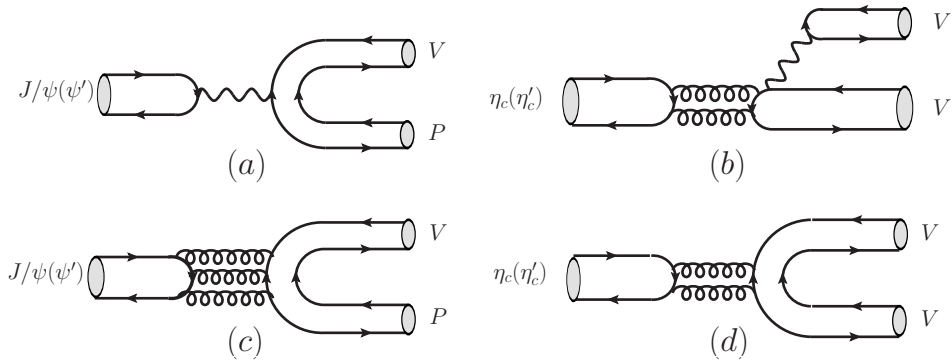}
 \caption{Schematic diagrams for the EM and strong
decays of $J/\psi (\psi^\prime)\to VP$ and $\eta_c(\eta_c^\prime)\to
VV$. }\label{fig-power}
\end{center}
\end{figure}

As the spin-0 partner of $J/\psi(\psi')$, the decays of $\eta_c
(\eta_c')\to VV$ provide an alternative way to examine the role
played by the intermediate meson loop (IML) effects in the
explanation of the ``$\rho\pi$ puzzle". In Ref.~\cite{Wang:2010iq},
due to the lack of experimental data at that moment, the
long-distance IML could not be well constrained, and it was assumed
that the IML contributed about $10\%$ of the amplitudes in
$\eta_c\to VV$ channel. With the recent availability of experimental
data from BES-III~\cite{BESIII:2011ac}, we can fit five channels of
$\eta_c\to VV$, i.e. $\omega\omega$, $\phi\phi$, $K^*\bar{K}^*$,
$\rho\rho$ and $\omega\phi$, to constrain both the short and
long-distance contributions. The short-distance contribution in
$\eta_c^\prime\to VV$ can thus be extracted by the ratio of the wave
functions at the origin between $\eta_c^\prime$ and $\eta_c$. At the
same time, the evolution of the IML to $\eta_c'$ gives the
long-distance contribution in $\eta_c^\prime\to VV$. Note that the
EM contributions via Fig.~\ref{fig-power}(b) is negligibly small in
comparison with the strong ones via Fig.~\ref{fig-power}(d). It
makes the decays of $\eta_c (\eta_c')\to VV$ an ideal place to
single out the roles played by the short and long-distance strong
transitions.

As follows, we will first provide the details of our formulations
for the short and long-distance IML transitions in Sec. II. The
numerical results and analysis are presented in Sec.~III, and a
brief summary is given in the last section.

\section{The model}

This section provides the details of our theoretical approach for
$\eta_c (\eta_c')\to VV$. The first part is a parametrization of the
short-distance contributions via hard gluon radiations in $\eta_c
(\eta_c^\prime)\to VV$ when the $c$ and $\bar{c}$ annihilate at the
wavefunction origin. The second part is the intermediated charmed
meson loop transitions which account for the long-distance
contributions. Namely, it describes non-negligible light quark pair
creations before the $c\bar{c}$ annihilation. These two mechanisms,
in principle, have no double counting in the decays.

\subsection{The parametrization for $\eta_c (\eta_c^\prime)\to VV$}

An obvious advantage for the exclusive decays of $\eta_c
(\eta_c^\prime)\to VV$ is that these transitions, similar to
$J/\psi(\psi^\prime)\to VP$, have only one unique Lorentz structure
for the $VVP$ couplings. As stressed a number of times before, this
will allow a parametrization of the effective coupling constant
contributed by different mechanisms. In Ref.~\cite{Zhao:2006cx}, a
parametrization scheme is proposed for the short-distance
transitions where the Okubo-Zweig-Iizuka singly disconnected (SOZI)
and doubly disconnected (DOZI) processes can be parameterized by the
gluon counting rule.

Following Ref.~\cite{Zhao:2006cx}, the transition amplitudes for
$\eta_c\to VV$ can be expressed as
\begin{eqnarray}
\label{isospin-1} \langle \phi\phi | \hat{V}_{gg}| \eta_c\rangle
&=& g R^2 (1+r)   \nonumber\\
\langle \omega\omega | \hat{V}_{gg}| \eta_c\rangle
&=& g(1+2r)  \nonumber\\
\langle \omega\phi | \hat{V}_{gg}| \eta_c\rangle &=& g r R \sqrt{2}
\nonumber\\
\langle K^{*+}K^{*-}|  \hat{V}_{gg}| \eta_c\rangle
&=& g R \nonumber\\
\langle \rho^+\rho^- | \hat{V}_{gg}| \eta_c\rangle &=& g  \ .
\end{eqnarray}
where $\hat{V}_{gg}$ is the $\eta_c\to gg\to (q\bar{q})(q\bar{q})$
potential, and parameter $g$ denotes the coupling strength of the
SOZI transitions.  Parameter $r$ is the ratio of the DOZI transition
over the SOZI transition. It should be pointed out that the
additional gluon exchanges in DOZI are not necessarily perturbative.
However, since they are higher twist contributions we expect that
their contributions would be small. This effect can be parameterized
by $r$, of which a small value suggests a suppressed DOZI
contribution~\cite{Zhao:2006cx}. We also introduce the SU(3) flavor
breaking parameter $R$, of which its deviation from unity reflects
the change of couplings due to the mass difference between $u/d$ and
$s$. The amplitudes for other charge combinations of $K^*\bar{K^*}$
and $\rho\rho$ are implicated.

A commonly used form factor is adopted in the calculation of the
partial decay widths:
\begin{equation}
{\cal F}^2({\bf p})=p^{2l}\exp(-{\bf p}^2/8\beta^2) \ ,
\end{equation}
where ${\bf p}$ and $l$ are the three momentum and relative angular
momentum of the final-state mesons, respectively, in the $\eta_c$
rest frame. We adopt $\beta=0.5$ GeV, which is the same as in
Refs.~\cite{close-amsler,close-kirk,close-zhao-f0,Zhao:2005im,Zhao:2007ze}.
Such a form factor will largely account for the size effects from
the spatial wavefunctions of the initial and final state mesons.

\subsection{Intermediate charmed meson loops}

\subsubsection{Formulation}

\begin{figure}[tb]
\begin{center}
\begin{tabular}{cc}
 \hspace{-3cm}\includegraphics[scale=1.0]{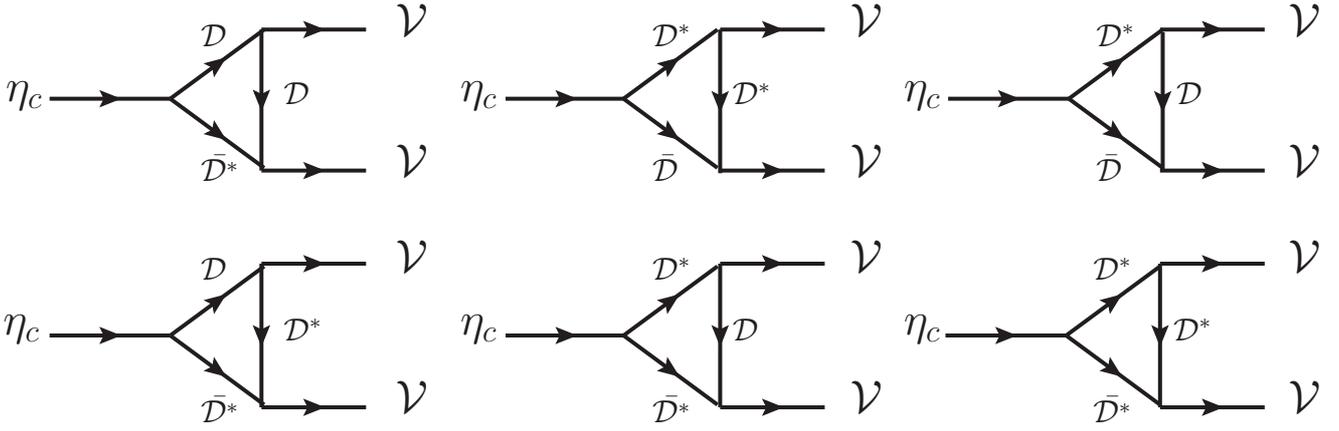}
\end{tabular}
\vspace{-20cm}\caption{Feyman diagrams for $\eta_c\to VV$ via
charmed meson loops. Here, $\mathcal{D}$ and $\mathcal{D}^*$
represent $(D^+,D^0,D_s^+)$ and $(D^{*+},D^{*0},D_s^{*+})$,
respectively, and $\mathcal{V}$ denotes the light vector meson.
}\label{fig-fey}
\end{center}
\end{figure}

A well-developed effective Lagrangian approach is applied to
estimate the long-distance IML transition
amplitudes~\cite{Liu:2009vv,Zhang:2009kr,Liu:2010um,Wang:2012mf}.
The Feynman diagrams for $\eta_c$ decays into $\rho\rho$,
$K^*\bar{K}^*$, $\omega\omega$, and $\phi\phi$ via the intermediate
charmed meson loops are illustrated in Fig.~\ref{fig-fey}. The
relevant effective Lagrangians are based on heavy quark symmetry
which describe the couplings between $S$-wave charmonium states and
charmed mesons~\cite{Colangelo:2003sa,Casalbuoni:1996pg} as the
following,
\begin{equation}
\mathcal{L}_2=i g_2 Tr[R_{c\bar{c}} \bar{H}_{2i}\gamma^\mu
{\stackrel{\leftrightarrow}{\partial}}_\mu \bar{H}_{1i}] + H.c.,
\end{equation}
where the $S$-wave charmonium states are expressed as
\begin{equation}
R_{c\bar{c}}=\left( \frac{1+ \rlap{/}{v} }{2} \right)\left(\psi^\mu
\gamma_\mu -\eta_c \gamma_5 \right )\left( \frac{1- \rlap{/}{v} }{2}
\right),
\end{equation}
and the charmed and anti-charmed meson triplet are
\begin{eqnarray}
H_{1i}&=&\left( \frac{1+ \rlap{/}{v} }{2} \right)[
\mathcal{D}_i^{*\mu}
\gamma_\mu -\mathcal{D}_i\gamma_5], \\
H_{2i}&=& [\bar{\mathcal{D}}_i^{*\mu} \gamma_\mu
-\bar{\mathcal{D}}_i\gamma_5]\left( \frac{1- \rlap{/}{v} }{2}
\right),
\end{eqnarray}
with $\mathcal{D}$ and $\mathcal{D}^*$ denote the pseudoscalar
($(D^{0},D^{+},D_s^{+})$) and vector charmed mesons
($(D^{*0},D^{*+},D_s^{*+})$), respectively. The Lagrangian
describing the interactions between light mesons and charmed mesons
reads
\begin{eqnarray}
 {\cal L} &=&\nonumber
 -ig_{\rho\pi\pi}\Big(\rho^+_\mu\pi^0{\stackrel{\leftrightarrow}{\partial}}{\!^\mu}\pi^-
 +\rho^-_\mu\pi^+{\stackrel{\leftrightarrow}{\partial}}{\!^\mu}\pi^0+\rho^0_\mu\pi^-{\stackrel{\leftrightarrow}{\partial}}{\!^\mu}\pi^+\Big)
  \\\nonumber
 &-& ig_{\mathcal{D}^*\mathcal{D}\mathcal{P}}(\mathcal{D}^i\partial^\mu \mathcal{P}_{ij}
 \mathcal{D}_\mu^{*j\dagger}-\mathcal{D}_\mu^{*i}\partial^\mu \mathcal{P}_{ij}\mathcal{D}^{j\dagger})
 +{1\over 2}g_{\mathcal{D}^*\mathcal{D}^*\mathcal{P}}
 \epsilon_{\mu\nu\alpha\beta}\,\mathcal{D}_i^{*\mu}\partial^\nu \mathcal{P}^{ij}
 {\stackrel{\leftrightarrow}{\partial}}{\!^\alpha} \mathcal{D}^{*\beta\dagger}_j
 \\\nonumber
 &-& ig_{\mathcal{D}\mathcal{D}\mathcal{V}} \mathcal{D}_i^\dagger {\stackrel{\leftrightarrow}{\partial}}{\!_\mu} \mathcal{D}^j(V^\mu)^i_j
 -2f_{\mathcal{D}^*\mathcal{D}\mathcal{V}} \epsilon_{\mu\nu\alpha\beta}
 (\partial^\mu \mathcal{V}^\nu)^i_j
 (\mathcal{D}_i^\dagger{\stackrel{\leftrightarrow}{\partial}}{\!^\alpha} \mathcal{D}^{*\beta j}-\mathcal{D}_i^{*\beta\dagger}{\stackrel{\leftrightarrow}{\partial}}{\!^\alpha} D^j)
 \\
 &+& ig_{\mathcal{D}^*\mathcal{D}^*\mathcal{V}} \mathcal{D}^{*\nu\dagger}_i {\stackrel{\leftrightarrow}{\partial}}{\!_\mu} \mathcal{D}^{*j}_\nu(\mathcal{V}^\mu)^i_j
 +4if_{\mathcal{D}^*\mathcal{D}^*\mathcal{V}} \mathcal{D}^{*\dagger}_{i\mu}(\partial^\mu \mathcal{V}^\nu-\partial^\nu
 \mathcal{V}^\mu)^i_j \mathcal{D}^{*j}_\nu,
 \label{eq:LDDV}
 \end{eqnarray}
with the convention $\epsilon^{0123}=+1$. In the above equation
$\mathcal{P}$ and $\mathcal{V}_\mu$ denote $3\times 3$ matrices for
the pseudoscalar octet and vector nonet,
respectively~\cite{Cheng:2004ru}, i.e.
\begin{eqnarray}
 \mathcal{P} &=& \left(\matrix{{\pi^0\over\sqrt{2}}+{\eta\over\sqrt{6}} & \pi^+ & K^+ \cr
 \pi^- & -{\pi^0\over\sqrt{2}}+{\eta\over\sqrt{6}} & K^0  \cr
 K^- & \bar{ K^0} & -\sqrt{2\over 3}\eta }\right), \ \
\mathcal{V} =
\left(\matrix{{\rho^0\over\sqrt{2}}+{\omega\over\sqrt{2}} & \rho^+ &
K^{*+} \cr
 \rho^- & -{\rho^0\over\sqrt{2}}+{\omega\over\sqrt{2}} & K^{*0}  \cr
 K^{*-} & \bar{ K^{*0}} & \phi }\right).
 \end{eqnarray}

The following kinematic conventions are adopted,  $\eta_c(p)\to
\mathcal{D}^{(*)}(p_1)\bar{\mathcal{D}}^{(*)}(p_3)[\mathcal{D}^{(*)}(p_2)]\to
V(k)V(q)$, where $\mathcal{D}^{(*)}$ in the square bracket denotes
the exchanged charmed meson in the triangle diagrams. The explicit
expressions of the amplitudes are
\begin{eqnarray}
\mathcal{M}_{\mathcal{D}\mathcal{D}^*[\mathcal{D}]}&=&\int\frac{d^4p_1}{(2\pi)^4}[-2g_{\eta_c\mathcal{D}\mathcal{D}^*}g_{\mathcal{D}\mathcal{D}\mathcal{V}}f_{\mathcal{D}\mathcal{D}^*\mathcal{V}}\epsilon_{\mu\nu\alpha\beta}(p_1+p_2)\cdot\epsilon_kq^\mu\epsilon_q^\nu
\nonumber\\
&\times&(p_2-p_3)^\alpha(p_1-p_3)_\lambda
(-g^{\lambda\beta}+\frac{p_3^\lambda p_3^\beta}{m_3^2})]\frac{1}{a_1
a_2 a_3}\mathcal{F}(p_i^2),\\\nonumber
\mathcal{M}_{\mathcal{D}\mathcal{D}^*[\mathcal{D}^*]}&=&\int\frac{d^4p_1}{(2\pi)^4}[-8g_{\eta_c\mathcal{D}\mathcal{D}^*}f_{\mathcal{D}\mathcal{D}^*\mathcal{V}}f_{\mathcal{D}^*\mathcal{D}^*\mathcal{V}}\epsilon_{\mu\nu\alpha\beta}k^\mu\epsilon_k^\nu(p_1+p_2)^\alpha
\\\nonumber&\times&(-g^{\beta\lambda}+\frac{p_2^\beta p_2^\lambda}{m_2^2})
(p_1-p_3)_\delta (-g^{\delta\theta}+\frac{p_3^\delta
p_3^\theta}{m_3^2})(\epsilon_{q\lambda}q_\theta-q_\lambda\epsilon_{q\theta})
\\\nonumber
&+&2g_{\eta_c\mathcal{D}\mathcal{D}^*}f_{\mathcal{D}\mathcal{D}^*\mathcal{V}}g_{\mathcal{D}^*\mathcal{D}^*\mathcal{V}}\epsilon_{\mu\nu\alpha\beta}k^\mu\epsilon_k^\nu(p_1+p_2)^\alpha(p_2-p_3)\cdot\epsilon_q
\\&\times&(p_1-p_3)_\delta(-g^{\beta\lambda}+\frac{p_2^\beta
p_2^\lambda}{m_2^2})(-g_\lambda^{\delta}+\frac{p_{3\lambda}
p_3^\delta}{m_3^2})]\frac{1}{a_1 a_2
a_3}\mathcal{F}(p_i^2),\\\nonumber
\mathcal{M}_{\mathcal{D}^*\mathcal{D}^*[\mathcal{D}]}&=&\int\frac{d^4p_1}{(2\pi)^4}[-4g_{\eta_c\mathcal{D}^*\mathcal{D}^*}f_{\mathcal{D}^*\mathcal{D}^*\mathcal{V}}^2\epsilon_{\mu\nu\alpha\beta}
\epsilon_{\rho\sigma\lambda\iota}\epsilon_{\eta\theta\xi\omega}p^\nu
p_1^\mu  (p_1+p_2)^\lambda \\&\times&k^\rho \epsilon_k^\sigma q^\eta
\epsilon_q^\theta (p_2-p_3)^\xi (-g^{\beta\iota}+\frac{p_1^\beta
p_1^\iota}{m_1^2})(-g^{\alpha\omega}+\frac{p_3^\alpha
p_3^\omega}{m_3^2})]\frac{1}{a_1 a_2 a_3}\mathcal{F}(p_i^2),\\
\mathcal{M}_{\mathcal{D}^*\mathcal{D}^*[\mathcal{D}^*]}&=&\int\frac{d^4p_1}{(2\pi)^4}[\mathcal{A}_1+\mathcal{A}_2+\mathcal{A}_3+\mathcal{A}_4]\frac{1}{a_1
a_2 a_3}\mathcal{F}(p_i^2), \nonumber
\end{eqnarray}
where
\begin{eqnarray}
 \mathcal{A}_1&\equiv &g_{\eta_c\mathcal{D}^*\mathcal{D}^*}g_{\mathcal{D}^*\mathcal{D}^*\mathcal{V}}^2\epsilon_{\mu\nu\alpha\beta}p^\nu p_1^\mu(p_1+p_2)\cdot\epsilon_k
 (p_2-p_3)\cdot \epsilon_q \\\nonumber
 &\times&(-g^{\beta\lambda}+\frac{p_1^\beta
 p_1^\lambda}{m_1^2})(-g_{\lambda\delta}+\frac{p_{2\lambda}
 p_{2\delta}}{m_2^2})(-g^{\delta\alpha}+\frac{p_3^\delta
 p_3^\alpha}{m_3^2}) \ ,\\
 \nonumber
 \mathcal{A}_2&\equiv &-4g_{\eta_c\mathcal{D}^*\mathcal{D}^*}g_{\mathcal{D}^*\mathcal{D}^*\mathcal{V}}f_{\mathcal{D}^*\mathcal{D}^*\mathcal{V}}\epsilon_{\mu\nu\alpha\beta}p^\nu
 p_1^\mu (p_1+p_2)\cdot \epsilon_k\\
 \nonumber&\times& (-g^{\beta\lambda}+\frac{p_1^\beta
 p_1^\lambda}{m_1^2})(-g_{\lambda\rho}+\frac{p_{2\lambda}
 p_{2\rho}}{m_2^2})(-g^{\alpha}_\delta+\frac{p_{3\delta}
 p_3^\alpha}{m_3^2})(\epsilon_q^\rho q^\delta-q^\rho
 \epsilon_q^\delta) \ ,\\
 \nonumber
 \mathcal{A}_3&\equiv &-4g_{\eta_c\mathcal{D}^*\mathcal{D}^*}g_{\mathcal{D}^*\mathcal{D}^*\mathcal{V}}f_{\mathcal{D}^*\mathcal{D}^*\mathcal{V}}\epsilon_{\mu\nu\alpha\beta}p^\nu
 p_1^\mu (-p_3+p_2)\cdot \epsilon_q \\
 \nonumber&\times&(-g^{\beta}_\lambda+\frac{p_1^\beta
 p_{1\lambda}}{m_1^2})(-g_{\sigma\rho}+\frac{p_{2\sigma}
 p_{2\rho}}{m_2^2})(-g^{\rho\alpha}+\frac{p_3^\rho
 p_3^\alpha}{m_3^2})(k^\sigma \epsilon_k^\lambda-\epsilon_k^\sigma k^\lambda) \ ,\\
 \nonumber
 \mathcal{A}_4&\equiv &16g_{\eta_c
 \mathcal{D}^*\mathcal{D}^*}f_{\mathcal{D}^*\mathcal{D}^*\mathcal{V}}^2\epsilon_{\mu\nu\alpha\beta}p^\nu p_1^\mu (-g^{\beta}_\lambda+\frac{p_1^\beta
 p_{1\lambda}}{m_1^2})(-g^{\alpha}_\rho+\frac{p_2^\alpha
 p_{2\rho}}{m_2^2})\\\nonumber&\times&(-g_{\theta\delta}+\frac{p_{3\theta}
 p_{3\delta}}{m_3^2})(q^\rho \epsilon_k^\lambda k^\theta \epsilon_q^\delta +
\epsilon_q^\rho k^\lambda q^\theta \epsilon_k^\delta
 -\epsilon_q^\rho \epsilon_k^\lambda q^\theta k^\delta - q^\rho k^\lambda \epsilon_k^\theta
 \epsilon_q^\delta) \ ,
\end{eqnarray}
with $a_1\equiv p_1^2-m_1^2, \ a_2\equiv p_2^2-m_2^2$, and
$a_3\equiv p_3^2-m_3^2$. The amplitudes
$\mathcal{M}_{\mathcal{D}^*\bar{\mathcal{D}}[\mathcal{D}]}$ and
$M_{\mathcal{D}^*\bar{\mathcal{D}}[\mathcal{D}^*]}$ have the same
expressions as
$\mathcal{M}_{\mathcal{D}\bar{\mathcal{D}}^*[\mathcal{D}]}$ and
$M_{\mathcal{D}\bar{\mathcal{D}}^*[\mathcal{D}^*]}$, respectively,
and are omitted for brevity. The full expressions of the transition
amplitudes for each $VV$ mode can be found in the Appendix.

Since the couplings in the effective Lagrangians are local ones,
ultra-violate divergence in the loop integrals is inevitable. We
introduce a tri-monopole form factor $\mathcal{F}(p_i^2)$
phenomenologically to take into account the non-local effects and
cut off the divergence in the loop integrals, i.e.
\begin{equation}
\mathcal{F}(p_i^2)=\prod\limits_i\left(
\frac{\Lambda_i^2-m_i^2}{\Lambda_i^2-p_i^2} \right ),
\end{equation}
where $m_i$ and $p_i$ are the mass and four momentum of the
corresponding exchanged particle, and the cut-off energy is chosen
as $\Lambda_i=m_i+\alpha\Lambda_{QCD}$ with $\Lambda_{QCD}=0.22$ GeV
\cite{Cheng:2004ru,Liu:2009vv,Liu:2010um}. The value of parameter
$\alpha$ will generally be determined by experimental data.

\subsubsection{Vertex couplings in the IML integrals}

Before proceeding to the numerical results, we first determine some
of the parameters included in this approach. In the chiral and heavy
quark limit, the following relations can be
obtained~\cite{Cheng:2004ru,Casalbuoni:1996pg},
\begin{eqnarray}
g_{\mathcal{D}\mathcal{D}\mathcal{V}}=g_{\mathcal{D}^*\mathcal{D}^*\mathcal{V}}=\frac{\beta
g_\mathcal{V}}{\sqrt{2}}, \
f_{\mathcal{D}\mathcal{D}^*\mathcal{V}}=\frac{f_{\mathcal{D}^*\mathcal{D}^*\mathcal{V}}}{m_{\mathcal{D}^*}}=\frac{\lambda
g_\mathcal{V}}{\sqrt{2}}, \
 g_\mathcal{V}=\frac{m_\rho}{f_\pi}, \\
g_{\eta_c \mathcal{D}\mathcal{D}^*}=g_{\eta_c \mathcal{D}^*
\mathcal{D}^*}\sqrt{\frac{m_\mathcal{D}}{m_{\mathcal{D}^*}}}m_{\eta_c}
=2g_2\sqrt{m_{\eta_c}m_{\mathcal{D}}m_{\mathcal{D}^*}},
\end{eqnarray}
where $\beta$ and $\lambda$ are commonly taken as $\beta=0.9, \
\lambda=0.56$ GeV$^{-1}$, while $f_\pi$ is the pion decay constant.

In principle, the coupling $g_2$ should be computed by
nonperturbative methods. If we simply estimate it with the vector
meson dominance (VMD) argument, it gives
$g_2=\sqrt{m_\psi}/(2m_\mathcal{D} f_\psi)$, where $m_\psi$ and
$f_\psi=405$ MeV being the mass and decay constant of $J/\psi$
\cite{Colangelo:2003sa}. This relation gives $g_{\eta_c \mathcal{D}
\mathcal{D}^*}=7.68$, which is a commonly adopted value in the
literature. For simplicity, we assume that $g_{\eta_c \mathcal{D}
\mathcal{D}^*}/g_{\eta_c^\prime \mathcal{D} \mathcal{D}^*}=1$ and it
is also applied to the $\eta_c(\eta_c^\prime)$ coupling to
$\mathcal{D}^*\mathcal{D}^*$.

It should be mentioned again that the decays
$\eta_c(\eta_c^\prime)\to VV$ are strong-interaction-dominant
processes with negligible EM contributions. They are different from
the decays $J/\psi (\psi^\prime) \to VP$ where the EM interaction
may still play an important role, especially in $\psi^\prime\to
VP$~\cite{Zhao:2006gw,Li:2007ky}. As shown in Fig.~\ref{fig-power},
a naive power counting indicates that
\begin{eqnarray}
\frac{\mathcal{M}_{em}}{\mathcal{M}_{strong}} &\sim&
\frac{\alpha_e}{\alpha_s^2}\ \
\mbox{for}\ \  J/\psi (\psi') \to VP, \nonumber\\
\frac{\mathcal{M}_{em}}{\mathcal{M}_{strong}} &\sim& \alpha_e\ \
\mbox{for}\ \ \eta_c (\eta_c')\to VV,
\end{eqnarray}
where ``$\mathcal{M}_{em}$" and ``$\mathcal{M}_{strong}$" denote the
leading EM and strong transition amplitudes, respectively. It
implies that the EM contribution in $\eta_c (\eta_c')\to VV$ will be
less important than that in $J/\psi (\psi') \to VP$. As a result,
the total (strong) amplitude of $\eta_c(\eta_c^\prime)\to VV$ can be
expressed as
\begin{eqnarray}
\mathcal{M}_{tot}(\eta_c(\eta_c^\prime))=\mathcal{M}_{short}+e^{i\theta(\theta')}\mathcal{M}_{long},
\end{eqnarray}
where $\mathcal{M}_{short}$ and $\mathcal{M}_{long}$ are the short
and long-distance amplitude, respectively, and $\theta \ (\theta')$
is the relative phase angle for the $\eta_c (\eta_c')$ decays. In
principle, $\theta (\theta')$ should be $0^\circ$ or $180^\circ$
which provide an arbitrary sign between those two real amplitudes.
Deviations from $0^\circ$ or $180^\circ$ imply that intermediate
light meson loops might have contributions. Note that the $VVP$
coupling has only one unique Lorentz structure. Any possible
contributions would be absorbed into the effective coupling
constant. By freeing $\theta (\theta')$, the imaginary amplitudes,
if necessary, would indirectly reflect the intermediate light meson
loops. We mention in advance that the data fitting for $\eta_c\to
VV$ favors that $\theta\simeq 180^\circ$, which indicates that the
light meson loops are rather small. This contribution can be better
studies given the availability of more precise measurement of the
branching ratios.


\section{Numerical results and analysis}

There are five explicit parameters to be determined in $\eta_c\to
VV$, among which three are from the short-distance transitions, i.e.
the SOZI strength $g$, the SU(3) breaking parameter $R$, and the
DOZI parameter $r$, while two are from the long-distance IML
transitions, i.e. the cut-off parameter $\alpha$ and the relative
angle $\theta$ between the short-distance and the long-distance
amplitudes. In contrast, other parameters in the previous section
have been determined independently and are treated as inputs.

It appears to be possible to constrain those parameters for
$\eta_c\to VV$ given the availability of the experimental data for
five decay channels. For the $\eta_c\to \omega\omega$ and
$\omega\phi$ channel, we use the half values of the up limits in our
fitting scheme. As follows, we first do a numerical fit of the data
and analyze the implicated constraints on the underlying dynamics.

In Table~\ref{tab-parameter} the fitted parameters are listed. The
following points can be learned from the fitted results:
\begin{itemize}
\item The SOZI strength, $g_{\eta_c}=(3.913\pm 1.03)\times 10^{-2}$,
  is larger than $g_{J/\psi}=1.75\times 10^{-2}$ determined in $J/\psi\to
  VP$~\cite{Wang:2012mf}. Since $c\bar{c}$ annihilate through three gluons in
  $J/\psi$ decays and  through two gluons in $\eta_c$ decays, this
  SOZI parameter satisfies approximately the relation established by the
  inclusive strong decay widths for $J/\psi\to 3g$ and $\eta_c\to
  2g$.

\item The SU(3) flavor symmetry breaking parameter $R$ is generally estimated
by   $f_\pi/f_K\sim 0.838$, with $f_\pi$ and $f_K$ the decay
constants of
  pion and kaon. As shown in Table~\ref{tab-parameter}, the fitted value $R=0.854\pm
  0.158$ is consistent with the decay constant ratio.

\item The DOZI parameter $r\sim -0.154$ is consistent with that
extracted in Refs.~\cite{Zhao:2006cx,Wang:2012mf}. It suggests that
the short-distance SOZI is dominant in $\eta_c\to VV$. In contrast
with $J/\psi\to VP$, the ambiguity due to the possible glueball
component inside the $\eta$ and $\eta'$ can be avoided.

\item It shows that there are large uncertainties with the relative phase between
the short and long-distance amplitudes.  This suggests that
contributions from the long-distance IML transitions are relatively
small in $\eta_c\to VV$. In comparison with $J/\psi\to VP$ the IML
contributions do not have much freedom since the vertex couplings in
these two processes are correlated in the limit of heavy quark
symmetry. It is thus natural to expect that the value of the form
factor parameter $\alpha$ is in a similar range as that determined
in $J/\psi\to VP$. In this work, by relaxing slightly the boundary
values for $\alpha$, i.e. with $\alpha= 0.30\pm 0.151$, we find that
the reduced $\chi^2$ can be improved significantly.

\end{itemize}

\begin{table}
\caption{The parameters in our model have been fitted in $\eta_c\to
VV$ processes. The reduced chi-square is also given in this table. }
\begin{tabular}{cc}
  \hline\hline
  parameter & value \\\hline
  $g_{\eta_c}$ & $(3.913\pm 1.03)\times 10^{-2}$ \\\hline
  $R$ & $0.854\pm 0.158$ \\\hline
   $r$& $-0.154\pm 0.117$ \\\hline
  $\theta$ & $174.29\pm 91.798$ \\\hline
  $\alpha$ & $0.30\pm 0.151$ \\\hline
  $\chi^2$ &3.67\\
  \hline\hline
\end{tabular}
\label{tab-parameter}
\end{table}

In Table~\ref{tab-etactovv} the fitted branching ratios are listed
and compared with the experimental data~\cite{Nakamura:2010zzi}. The
central value of the total decay width $\Gamma_{\eta_c}=28.6\pm 2.2
\ \mathrm{MeV}$ is adopted. The branching ratios from the exclusive
short and long-distance transitions are also listed. It shows that
the short-distance transitions are dominant in $\eta_c\to VV$ as
expected, while the IML contributions are rather small since the
$\eta_c$ mass is far away from the open charm threshold.

With the above determined parameters for the short and long-distance
transition amplitudes in $\eta_c\to VV$, we now extend the
calculation to $\eta_c'\to VV$.

As discussed earlier, the short-distance decay amplitudes for
$\eta_c$ and $\eta_c'$ are proportional to their wavefunctions at
the origins. They are spin-0 partner of $J/\psi$ and $\psi'$,
respectively. Therefore, the ratio of the wavefunctions at their
origins can be related to that for $J/\psi$ and $\psi'$, i.e.
$|\eta_c^\prime(0)|/|\eta_c(0)|\simeq
|\psi_{2S}(0)|/|\psi_{1S}(0)|$. It has been broadly discussed in the
literature~\cite{Eichten:2004uh,Eichten:2005ga,Eichten:1978tg} that
$|\psi_{2S}(0)|/|\psi_{1S}(0)|$ can be extracted from either the
mass splitting between the $^1S_0$ and $^3S_1$ state or the lepton
pair decay widths of $\psi_{2S}$ and $\psi_{1S}$. For instance, the
estimate of Ref.~\cite{Eichten:1978tg} gives
$|\psi_{2S}(0)|^2/|\psi_{1S}(0)|^2=0.62\pm0.16$. By adopting
$|\eta_c^\prime(0)|/|\eta_c(0)|\simeq
|\psi_{2S}(0)|/|\psi_{1S}(0)|$, we can determine the magnitude of
the short-distance amplitudes for $\eta_c'\to VV$ via the $\eta_c$
decays.

For the long-distance transition amplitude, the form factor
parameter $\alpha_{\eta_c'}$ cannot be well constrained. However,
one can determine its upper limit based on the properties of the
meson loop integrals. In comparison with the IML in $\eta_c\to VV$,
the IML integrals in $\eta_c'\to VV$ involves the change of the
first vertex couplings, i.e. $\eta_c'\mathcal{D}\mathcal{D}^*$ and
$\eta_c'\mathcal{D}^*\mathcal{D}^*$, and the initial mass change
from $\eta_c$ to $\eta_c'$. Apart from the vertex couplings, the
initial mass change will lead to enhanced IML contributions if the
same form factor parameter $\alpha_{\eta_c'}=\alpha_{\eta_c}$ is
adopted. Due to this feature, we will also investigate the
sensitivities of the predictions within
$\alpha_{\eta_c'}=\alpha_{\eta_c}=0.30\pm 0.151$. This should be a
reasonable range for the estimate of the upper limit of the
long-distance IML contributions with an uncertainty.

The last parameter, the relative phase angle $\theta'$ between the
short and long-distance amplitudes, is a free one in this
formulation. Taking into account the uncertainties with the values
of $|\eta_c^\prime(0)|$ and $\alpha_{\eta_c'}$, we study two
situations of $\theta'$-dependence of the branching ratios for
$\eta_c'\to VV$ as follows.

\begin{figure}
\begin{center}
\hspace{-0.5cm}
\includegraphics[scale=0.65]{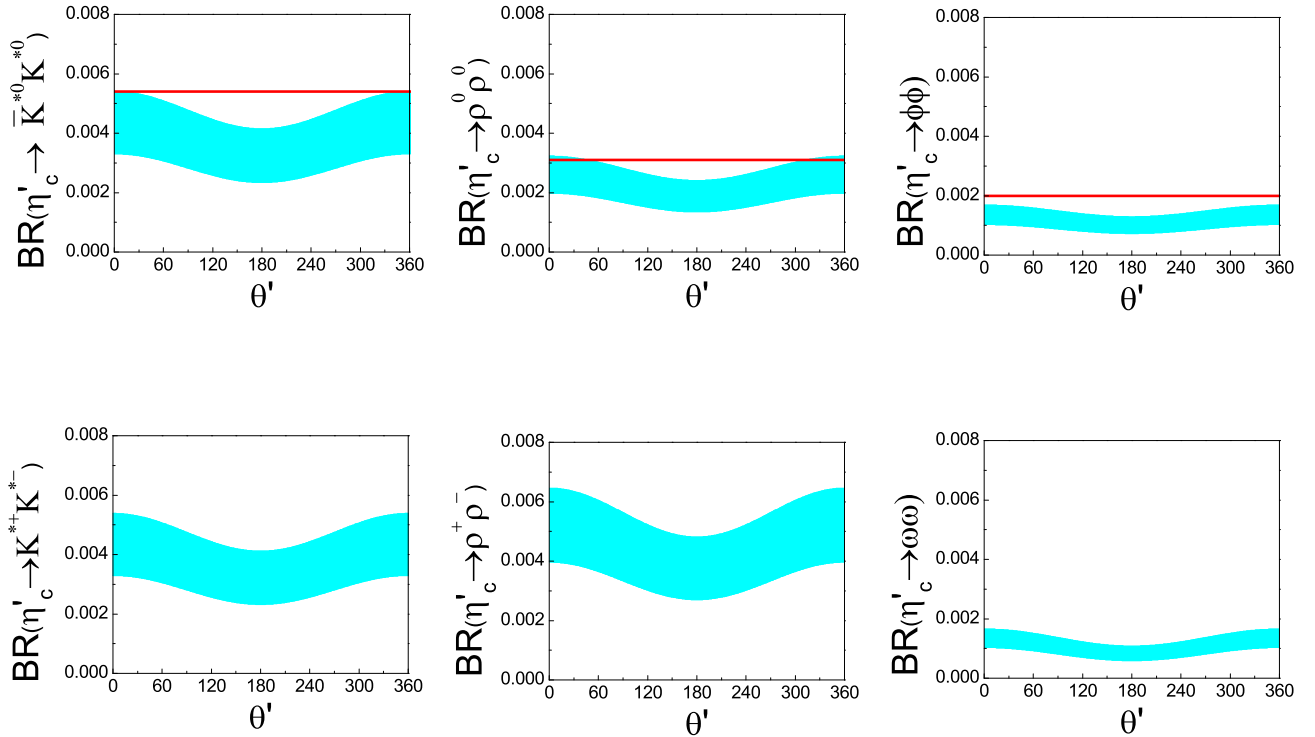}
\vspace{-3cm} \caption{The $\theta'$-dependence of the branching
ratios of $\eta_c^\prime\to VV$ with
$\alpha_{\eta_c'}=\alpha_{\eta_c}=0.30$. The shadowed bands reflect
the uncertainties arising from
$|\eta_c^\prime(0)|/|\eta_c(0)|=0.62\pm 0.16$, while the red solid
lines indicate the upper limit from the experimental
measurement~\cite{BESIII:2011ac}. }\label{fig-etacprime}
\end{center}
\end{figure}

First, in Fig.~\ref{fig-etacprime} we fix
$\alpha_{\eta_c'}=\alpha_{\eta_c}=0.30$ and explore the
$\theta'$-dependence of the branching ratios. The shadowed bands are
the range given by the uncertainties due to the value of the
wavefunction at the origin, i.e.
$|\eta_c^\prime(0)|/|\eta_c(0)|\simeq
|\psi_{2S}(0)|/|\psi_{1S}(0)|=0.62\pm 0.16$. The solid lines are the
experimental upper limits from the BES-III
experiment~\cite{BESIII:2011ac}. It shows that for those measured
channels, i.e. $\eta_c^\prime\to K^{*0}\bar{K}^{*0}$,
$\rho^0\rho^0$, and $\phi\phi$, the predicted branching ratios are
consistent with the experimental limits. In fact, the contributions
from the long-distance IML transitions are still relatively smaller
than those from the short-distance ones. Therefore, the
interferences at different relative phase angles cause only about
20\% deviations.

\begin{figure}
\begin{center}
\hspace{-0.5cm}
\includegraphics[scale=0.65]{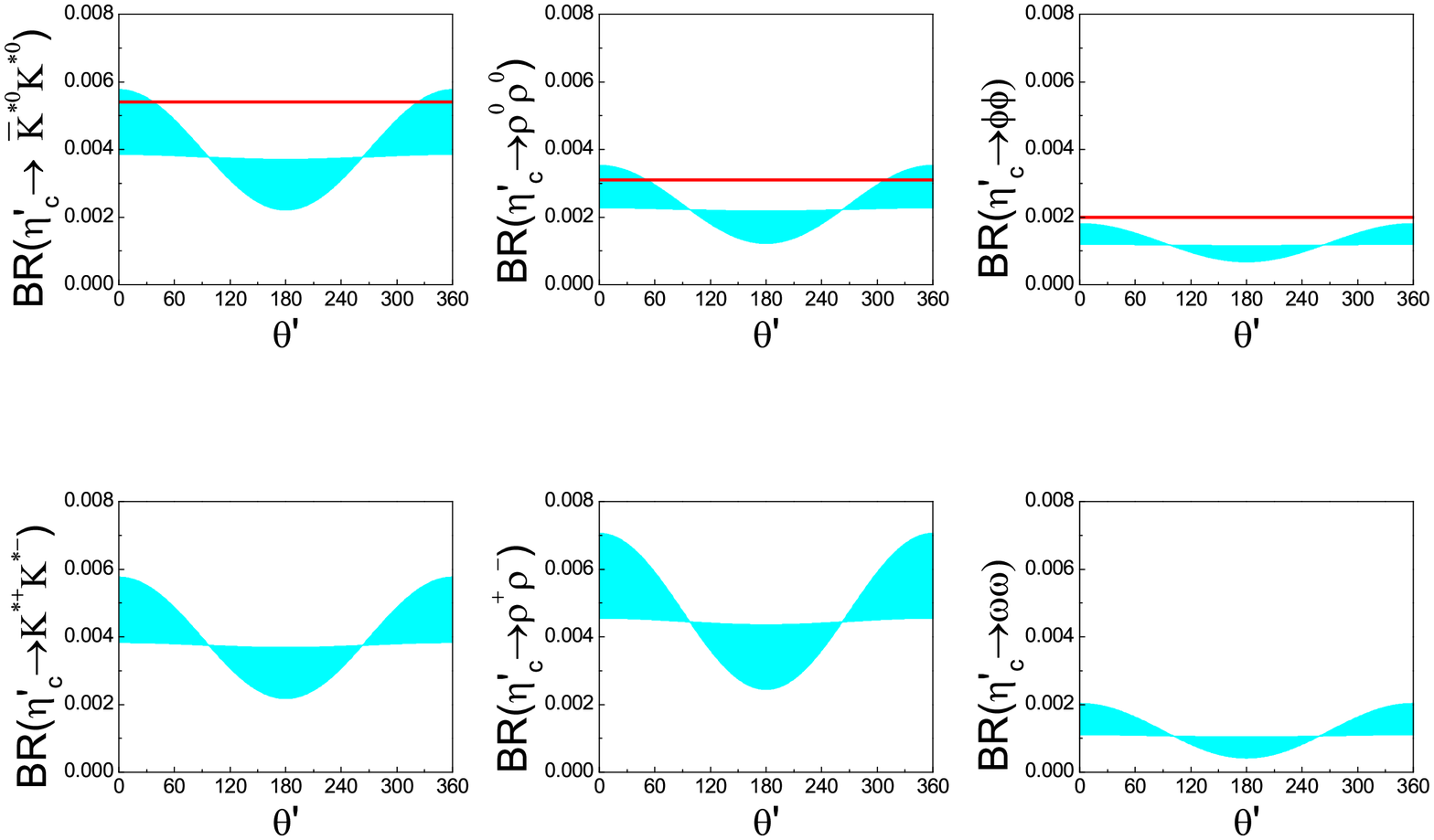}
\vspace{-3cm} \caption{The $\theta'$-dependence of the branching
ratios of $\eta_c^\prime\to VV$ with
$|\eta_c^\prime(0)|/|\eta_c(0)|=0.62$. The shadowed areas reflect
the uncertainties arising from $\alpha_{\eta_c'}=0.30\pm 0.151$,
while the red solid lines are the same as in
Fig.~\ref{fig-etacprime}. }\label{fig-etacprime-2}
\end{center}
\end{figure}

Secondly, in Fig.~\ref{fig-etacprime-2}, we fix the central value
$|\eta_c^\prime(0)|/|\eta_c(0)|=0.62$ and explore the
$\theta'$-dependence of the branching ratios within the range of
$\alpha_{\eta_c'}=0.30\pm 0.151$. By comparing the values at
$\theta'=0^\circ$ in Figs.~\ref{fig-etacprime} and
~\ref{fig-etacprime-2}, we can see that the uncertainties with
$|\eta_c^\prime(0)|$ and $\alpha_{\eta_c'}$ cause similar magnitude
of uncertainties with the predicted branching ratios. Since both the
short and long-distance amplitudes are calculated as real
quantities, the nodes structures at $\theta'=90^\circ$ and
$270^\circ$ in Fig.~\ref{fig-etacprime-2} locate the phase angles
that the IML transitions only contribute to the imaginary part.
Since the magnitude is small, its changes to the predicted branching
ratios are not significant. In contrast, at $\theta'=0^\circ$ or
$180^\circ$, the IML transitions contribute to the real part with
different signs. Its interferences are amplified by the
short-distance amplitude, thus lead to much larger effects to the
predictions as shown by the shadowed areas.

The following points can be further learned in order:
\begin{itemize}
\item
In $\eta_c'\to VV$ the relative strength of the IML amplitudes to
the short-distance ones is not as large as that in $\psi'\to VP$.
This should be understandable due to the different open charm
thresholds in $\psi'\to VP$ and $\eta_c'\to VV$. For the latter
process, the first open charm threshold is $D^* \bar{D}+c.c.$, while
for $\psi'\to VP$ it is $D\bar{D}$. As a consequence, the dominance
of the short-distance contributions can make the relations by
Eqs.~(\ref{rela-1}) and (\ref{rela-2}) to be satisfied. It is
unlikely to have a drastic variation of the ratios as observed in
$J/\psi$ and $\psi'\to VP$~\cite{Wang:2012mf}.

\item Since the long-distance IML transitions do not contribute to the
$\omega\phi$ decay mode at leading order, it is interesting to
recognize that a precise measurement of $BR(\eta_c(\eta_c^\prime)\to
\omega\phi)$ would provide a strong constraint on the short-distance
process. In particular, it will constrain the DOZI transition
parameter $r$ in $\eta_c'\to VV$. By adopting the value $r=-0.154\pm
0.117$ determined in $\eta_c\to VV$, we predict $BR(\eta_c^\prime\to
\omega\phi)=(1.82\pm 0.47)\times 10^{-4}$, which can be tested by
future BES-III data.

\end{itemize}

Combining these two decays together, i.e. $\eta_c(\eta_c^\prime)\to
VV$ and $J/\psi(\psi^\prime)\to VP$~\cite{Wang:2012mf}, we find that
the IML transitions do not contribute strongly in $\eta_c'\to VV$
compared with those in $\psi'\to VP$ because of the difference of
open charm thresholds. In contrast, the IML contributions in both
$\eta_c\to VV$ and $J/\psi\to VP$ are rather small. As discussed
before that the EM contribution can be neglected in
$\eta_c(\eta_c^\prime)\to VV$, the relatively small contribution
from the long-distance IML implies that Eq.(\ref{rela-1}) would be
well respected. We stress that our investigation suggests that the
same IML mechanism would have rather different manifestations in
different processes. Therefore, a systematic study of those
correlated processes is essential for a better understanding of the
HSR violation and disentangling the long-standing ``$\rho\pi$
puzzle".

\begin{table}
\caption{Branching ratios of $\eta_c\to VV$ in comparison with the
experimental data~\cite{Nakamura:2010zzi}. The exclusive branching
ratios from the short and long-distance transitions are also
included. The total width $\Gamma_{\eta_c}=28.6 \ \mathrm{MeV}$ is
adopted.}
\begin{tabular}{ccccc}
  \hline\hline
  $BR(\eta_c\to VV)$ & short-distance & long-distance & tot. & exp. \\
  \hline
  $\omega\omega$ & $1.92\times 10^{-3}$ & $3.2\times 10^{-6}$ & $1.76\times 10^{-3}$ & $<3.1\times 10^{-3}$
  \\\hline
  $\phi\phi$ & $2.1\times 10^{-3}$ & $2.33\times 10^{-6}$ & $1.96\times 10^{-3}$ & $(2.7\pm0.9)\times 10^{-3}$
  \\\hline
  $K^*\bar{K}^*$ & $1.34\times 10^{-2}$ & $1.09\times 10^{-5}$ & $1.27\times 10^{-2}$ & $(9.2\pm3.4)\times 10^{-3}$
  \\\hline
  $\rho\rho$ & $1.19\times 10^{-2}$ & $9.53\times 10^{-6}$ & $1.13\times 10^{-2}$ & $(2.0\pm 0.7)\times 10^{-2}$
  \\\hline
  $\omega\phi$ & $3.25\times 10^{-4}$ & 0 & $3.25\times 10^{-4}$ & $<1.7\times 10^{-3}$ \\
  \hline\hline
\end{tabular}
\label{tab-etactovv}
\end{table}

\section{Summary}

In this work, we have studied the decay properties of $\eta_c$ and
$\eta_c'\to VV$ as an alternative process for the test of
intermediate meson loop transitions. The availability of
experimental data for $\eta_c\to VV$ allows the determination of our
model parameters which then can be used as inputs in
$\eta_c^\prime\to VV$. Nevertheless, we show that the correlation
between $\eta_c (\eta_c')\to VV$ and $J/\psi (\psi')\to VP$ is
essentially important for isolating the short and long-distance
transition amplitudes in the $\eta_c$ and $\eta_c'$ decays. The
predicted branching ratios for $\eta_c'\to VV$ turn out to be
consistent with the recent data from BES-III~\cite{BESIII:2011ac}.

This investigation provides additional information for the IML
transitions. Similar to $J/\psi\to VP$, the IML effects are
negligible in $\eta_c\to VV$ since their masses are far below the
open charm thresholds. It was found that the IML transitions played
an important role in $\psi'\to VP$ since the mass of $\psi'$ was
close to the $D\bar{D}$ threshold~\cite{Wang:2012mf}. In contrast,
the IML contributions in $\eta_c'\to VV$ are not as significant as
those in $\psi'\to VP$ since the contributing open charm threshold
starts with $D^*\bar{D}+c.c.$ which is much higher than $D\bar{D}$.
This interesting feature suggests  that the same IML mechanism would
have rather different manifestations in different processes. We
stress that such a correlated study is crucial for disentangling
some of those long-standing puzzles in the charmonium energy region.

We should also mention that our conclusions are based on the
hypothesis that the flavor components of $\eta_c$ and $\eta_c'$ are
dominated by $c\bar{c}$. Thus, the connection of their spatial
wavefunctions with those of $J/\psi$ and $\psi'$ will make sense. If
$\eta_c$ or $\eta_c'$ possesses some other internal
structures~\cite{Feldmann:2000hs,Anselmino:1993yg,Anselmino:1990vs,Tsai:2011dp},
the relation between their branching ratios will be affected to some
extent, which however, is not our focus in this work.

\section*{Acknowledgement}

This work is supported, in part, by National Natural Science
Foundation of China (Grant No. 11035006), Chinese Academy of
Sciences (KJCX2-EW-N01), and Ministry of Science and Technology of
China (2009CB825200).

\section*{Appendix}

The IML transition amplitudes for the $\eta_c$ decays into
$\rho\rho$, $K^*\bar{K}^*$, $\omega\omega$, and $\phi\phi$ as
illustrated by Fig.~\ref{fig-fey} are presented here for a
reference:
\begin{eqnarray}
\mathcal{M}(\eta_c\to\rho^0\rho^0)&=&\mathcal{M}_{D^+D^{*-}[D^+]}
+\mathcal{M}_{D^{*+}D^{-}[D^{*+}]}
+\mathcal{M}_{D^{*+}D^{-}[D^+]}\\\nonumber
&+&\mathcal{M}_{D^+D^{*-}[D^{*+}]} +\mathcal{M}_{D^{*+}D^{*-}[D^+]}
+\mathcal{M}_{D^{*+}D^{*-}[D^{*+}]}\\\nonumber
&+&\mathcal{M}_{D^0\bar{D}^{*0}[D^0]}
+\mathcal{M}_{D^{*0}\bar{D}^{0}[D^{*0}]}
+\mathcal{M}_{D^{*0}\bar{D}^{0}[D^0]}\\\nonumber
&+&\mathcal{M}_{D^0\bar{D}^{*0}[D^{*0}]}
+\mathcal{M}_{D^{*0}\bar{D}^{*0}[D^0]}
+\mathcal{M}_{D^{*0}\bar{D}^{*0}[D^{*0}]}+c.c.,\\
\mathcal{M}(\eta_c\to\rho^+\rho^-)&=&\mathcal{M}_{D^+D^{*-}[D^0]}
+\mathcal{M}_{D^{*+}D^{-}[D^{*0}]}
+\mathcal{M}_{D^{*+}D^{-}[D^0]}\\\nonumber
&+&\mathcal{M}_{D^+D^{*-}[D^{*0}]} +\mathcal{M}_{D^{*+}D^{*-}[D^0]}
+\mathcal{M}_{D^{*+}D^{*-}[D^{*0}]}\\\nonumber
&+&\mathcal{M}_{D^0\bar{D}^{*0}[D^-]}
+\mathcal{M}_{D^{*0}\bar{D}^{0}[D^{*-}]}
+\mathcal{M}_{D^{*0}\bar{D}^{0}[D^-]}\\\nonumber
&+&\mathcal{M}_{D^0\bar{D}^{*0}[D^{*-}]}
+\mathcal{M}_{D^{*0}\bar{D}^{*0}[D^-]}
+\mathcal{M}_{D^{*0}\bar{D}^{*0}[D^{*-}]}+c.c.,\\
\mathcal{M}(\eta_c\to\omega\omega)&=&\mathcal{M}(\eta_c\to\rho^0\rho^0),\\
\mathcal{M}(\eta_c\to
K^{*0}\bar{K}^{*0})&=&\mathcal{M}_{D_s^+D_s^{*-}[D^+]}
+\mathcal{M}_{D_s^{*+}D_s^{-}[D^{*+}]}
+\mathcal{M}_{D_s^{*+}D_s^{-}[D^+]}\\\nonumber
&+&\mathcal{M}_{D_s^+D_s^{*-}[D^{*+}]}
+\mathcal{M}_{D_s^{*+}D_s^{*-}[D^+]}
+\mathcal{M}_{D_s^{*+}D_s^{*-}[D^{*+}]}\\\nonumber
&+&\mathcal{M}_{D^-D^{*+}[D_s^-]}
+\mathcal{M}_{D^{*-}D^{+}[D_s^{*-}]}
+\mathcal{M}_{D^{*-}D^{+}[D_s^-]}\\\nonumber
&+&\mathcal{M}_{D^-D^{*+}[D_s^{*-}]}
+\mathcal{M}_{D^{*-}D^{*+}[D_s^-]}
+\mathcal{M}_{D^{*-}D^{*+}[D_s^{*-}]}+c.c.,\\
\mathcal{M}(\eta_c\to
K^{*+}K^{*-})&=&\mathcal{M}_{D_s^+D_s^{*-}[D^0]}
+\mathcal{M}_{D_s^{*+}D_s^{-}[D^{*0}]}
+\mathcal{M}_{D_s^{*+}D_s^{-}[D^0]}\\\nonumber
&+&\mathcal{M}_{D_s^+D_s^{*-}[D^{*0}]}
+\mathcal{M}_{D_s^{*+}D_s^{*-}[D^0]}
+\mathcal{M}_{D_s^{*+}D_s^{*-}[D^{*0}]}\\\nonumber
&+&\mathcal{M}_{D^0\bar{D}^{*0}[D_s^-]}
+\mathcal{M}_{D^{*0}\bar{D}^{0}[D_s^{*-}]}
+\mathcal{M}_{D^{*0}\bar{D}^{0}[D_s^-]}\\\nonumber
&+&\mathcal{M}_{D^0\bar{D}^{*0}[D_s^{*-}]}
+\mathcal{M}_{D^{*0}\bar{D}^{*0}[D_s^-]}
+\mathcal{M}_{D^{*0}\bar{D}^{*0}[D_s^{*-}]}+c.c.,\\
\mathcal{M}(\eta_c\to\phi\phi)&=&\mathcal{M}_{D_s^+D_s^{*-}[D_s^+]}
+\mathcal{M}_{D_s^{*+}D_s^{-}[D_s^{*+}]}
+\mathcal{M}_{D_s^{*+}D_s^{-}[D_s^+]}\\\nonumber
&+&\mathcal{M}_{D_s^+D_s^{*-}[D_s^{*+}]}
+\mathcal{M}_{D_s^{*+}D_s^{*-}[D_s^+]}
+\mathcal{M}_{D_s^{*+}D_s^{*-}[D_s^{*+}]}\\\nonumber
&+&\mathcal{M}_{D_s^-D_s^{*+}[D_s^-]}
+\mathcal{M}_{D_s^{*-}D_s^{+}[D_s^{*-}]}
+\mathcal{M}_{D_s^{*-}D_s^{+}[D_s^-]}\\\nonumber
&+&\mathcal{M}_{D_s^-D_s^{*+}[D_s^{*-}]}
+\mathcal{M}_{D_s^{*-}D_s^{*+}[D_s^-]}
+\mathcal{M}_{D_s^{*-}D_s^{*+}[D_s^{*-}]}+c.c.\\\nonumber
\end{eqnarray}

\end{document}